\documentclass[usenatbib,referee]{ctextemp_mn2e}

\title[the minimum mass ratio of W UMa binaries]
{On the minimum mass ratio of W UMa binaries}

\author[Jiang et al.]{Dengkai Jiang$^{1,2,3}$\thanks{E-mail:
dengkai@ynao.ac.cn}, Zhanwen Han$^{1,3}$, Jiancheng Wang$^{1,3}$,
Tianyu Jiang$^{1,2,3}$ \and and Lifang Li$^{1,3}$\\
$^{1}$National Astronomical Observatories, Yunnan Observatory,
Chinese Academy of Sciences, P.O. Box 110,
Kunming,\\
\ \ \ \  \ \ \ \ \ \ \ \ \ \ \ \ \ \ \ \ \ \ \ \ \ \ \ \ \ \ \ \ \
\ \ \ \  Yunnan Province, 650011, P.R. China\\
$^{2}$Graduate University of Chinese Academy Sciences, Beijing,
100039, P.R. China\\
$^{3}$Key Laboratory for the Structure and Evolution of Celestial
Objects, Chinese Academy of Sciences, Kunming 650011, P.R. China}
\begin{document}
\input ctextemp_psfig.sty
\date{Accepted .... Received .....; in original form ....}

\pagerange{\pageref{firstpage}--\pageref{lastpage}} \pubyear{2009}

\maketitle

\label{firstpage}

\begin{abstract}
Using Eggleton's stellar evolution code, we study the minimum mass
ratio ($q_{\rm min}$) of W Ursae Majoris (W UMa) binaries that have
different primary masses. It is found that the minimum mass ratio of
W UMa binaries decreases with increasing mass of the primary if the
primary's mass is less than about 1.3$M_{\rm \odot}$, and above this
mass the ratio is roughly constant. By comparing the theoretical
minimum mass ratio with the observational data, it is found that the
existence of low-$q$ systems can be explained by the different
structure of the primaries with different masses. This suggests that
the dimensionless gyration radius ($k_1^2$) and thus the structure
of the primary is very important in determining the minimum mass
ratio. In addition, we investigate the mass loss during the merging
process of W UMa systems and calculate the rotation velocities of
the single stars formed by the merger of W UMa binaries due to tidal
instability. It is found that in the case of the conservation of
mass and angular momentum, the merged single stars rotate with a
equatorial velocity of about $588\sim819$ km s$^{-1}$, which is much
larger than their break-up velocities ($v_{\rm b}$). This suggests
that the merged stars should extend to a very large radius
(3.7$\sim$5.3 times the radii of the primaries) or W UMa systems
would lose a large amount of mass (21$\sim$33 per cent of the total
mass) during the merging process. If the effect of magnetic braking
is considered, the mass loss decreases to be 12$\sim$18 per cent of
their total masses. This implies that the significant angular
momentum and mass might be lost from W UMa systems in the course of
the merging process, and this kind of mass and angular momentum loss
might be driven by the release of orbital energy of the secondaries,
which is similar to common-envelope evolution.
\end{abstract}


\section{Introduction}
W Ursae Majoris (W UMa) binaries are eclipsing variables in which
two components are in contact or overflowing their Roche limiting
surfaces. The components of W UMa binaries share a common convective
envelope. In general, W UMa systems have total system masses
$1M_{\rm \odot}\leq(M_1+M_2)\leq3M_{\rm \odot}$ and orbital periods
between 0.22 and 1 days \citep{Gazeas06}. They are very common
systems and can be discovered in field, open clusters, and globular
clusters \citep{Kaluzny 1993, Rucinski 1994, Rucinski 1998,ruc00}.
There is at least one W UMa binary for every 500 main sequence stars
in the solar neighborhood \citep{Rucinski 2002, Rucinski 2006}.
\citet{Rucinski 1994} gave the relative frequency of occurrence of
one W UMa system per $275 \pm 75$ ordinary dwarfs in open clusters.
In globular clusters, the relative frequency of occurrence of W UMa
systems was also found to be very high \citep{ruc00}.

\citet{Eggleton 2006} pointed out that many binaries of short period
can be expected to evolve into contact, and there is only a small
region in the initial orbital period and mass ratio plane where it
does not contact if the Roche-lobe overflow starts while the primary
is still in the main sequence band. W UMa systems have the least
amounts of angular momentum that binaries made of main-sequence
components can have, so they are important sources for testing the
angular momentum evolution of binaries \citep{ruc00, Selam 2004}.
More importantly, W UMa binaries can be used to study Galactic
structure because they have high spatial frequency of occurrence,
ease of detection and provide an absolute magnitude
calibration\citep{Rucinski 1997}. In addition, W UMa systems are the
possible progenitors of blue stragglers/FK Comae-type (FK Com) stars
\citep{Qian 2005}. Some blue stragglers in star clusters are likely
formed from the merger of W UMa binaries \citep{Lombardi et al
2002}. The investigation of the merger of W UMa binaries can help us
to understand the formation theory on blue stragglers and FK
Com-type stars.

Theoretical studies indicate that W UMa system would merge into a
fast-rotating single star due to the tidal instability (i.e.
Darwin's instability) when the spin angular momentum of the system
is more than a third of its orbital angular momentum \citep{Hut
1980, Eggleton 2001}. The occurrence of the tidal instability in W
UMa binaries determines the minimum mass ratio ($q_{min}$) of these
systems. W UMa binaries with mass ratio $q =M_2/M_1 \leq q_{min}$
should not be observed since they have merged into fast-rotating
stars within a tidal timescale (about $10^3-10^4$ yr). Therefore,
the minimum mass ratio is a very important parameter in
investigating the evolution and the merger of W UMa systems.

The minimum mass ratio of W UMa binaries has been investigated by
many authors \citep{Rasio 1995, Li 2004, Li 2005, Arbutina 2007,
Arbutina 2009}. If the rotation of the secondaries in W UMa systems
is neglected, the minimum mass ratio of W UMa-type systems is
derived to be of about 0.09 \citep{Rasio 1995}. If the rotation of
the secondaries is taken into account and $k_1^2=k_2^2=0.06$ ( where
$k_1^2=I_1/(M_1R_1^2), k_2^2=I_2/(M_2R_2^2)$ are the dimensionless
gyration radii for the primary and the secondary), the cutoff mass
ratio of W UMa systems is derived to be of 0.076 \citep{Li 2006}. If
it is assumed that the primary is radiative main-sequence star
($k_1^2\approx0.075$) and the secondary is fully convective star
($k_2^2\approx0.205$), the theoretical minimum mass ratio is derived
to be of about 0.094-0.109 \citep{Arbutina 2007}. These results
predict that W UMa binaries with a mass ratio less than the minimum
mass ratio should not be observed. However, the mass ratios of some
observed W UMa binaries are smaller than the theoretical minimum
mass ratio, such as V857 Her \citep[$q$=0.065][]{Qian 2005}, AW UMa
\citep[$q$=0.075][]{Rucinski 1992} and SX Crv
\citep[$q$=0.079][]{Zola 2004}. This can be explained by W UMa
systems with slightly evolved primaries or differential rotation of
their components \citep{Rasio 1995, Li 2006, Arbutina 2007}. In
order to remove the difference between observations and theoretical
predictions, \citet{Arbutina 2009} obtained a theoretical minimum
mass ratio of about 0.070-0.074 if the effects of rotation which
increase the central concentration are included. However, little has
been done to investigate the minimum ratio of W UMa systems that
have different primary masses. \citet{Hurley 2000} showed that there
is a difference in the structure of the main-sequence stars with
different masses. Because the dimensionless gyration radius of the
primary depends on the structure \citep{Li 2005}, this may introduce
the structure of the primaries with different masses that may need
to be taken into account in determining the minimum mass ratio of W
UMa systems.

W UMa systems with $q \leq q_{min}$ are unstable and undergo rapid
merging, which might result in the formation of the rapidly rotating
single stars \citep[blue stragglers or FK Com,][]{Webbink 1976,
ste95, Li 2004, Li 2005}. During the merger of a W UMa system, the
secondary enters the primary and it, together with the primary would
spin up. Therefore, the equatorial velocity of the single star
formed by the merger of W UMa binary would be larger than that of W
UMa binaries which is of 100-200 km s$^{-1}$ \citep{Selam 2004}.
However, the rotational velocities of the blue stragglers in field
are found to be normal \citep{Carney 1981}, while blue stragglers in
M67 are rotating slowly compared with main-sequence stars
\citep{Mathys 1991}. \citet{De Marco 2005} measured the rotation
velocities of five rapidly-rotating blue stragglers, $v \, \sin i$
of about 120, 100, 225, 50, and 50km s$^{-1}$, but this information
cannot constrain their origin as stellar collision or binary merger
because of the lack of clear theoretical predictions. For FK Com ($v
\, \sin i$ $\sim$ 160 km s$^{-1}$), its angular momentum is about 3
times smaller than that of the orbital motion in a typical W UMa
binary \citep{Rucinski 1990}. Therefore, a large amount of mass and
angular momentum must be lost from W UMa systems if some blue
stragglers or FK Com stars are formed from the merger of W UMa
systems. However, it is uncertain how much of the mass is lost
during the merging process.

The purpose of this paper is to study the minimum mass ratio of W
UMa systems and the mass loss during the merging process. Employing
Eggleton's stellar evolution code, we study the structure of the
primary with different mass, and then determine the minimum mass
ratio of W UMa binaries. We compare the theoretical minimum mass
ratio with the observational data and find that the existence of
low-q systems can be explained by the different structure of the
primaries with different masses. We find that the structure of the
primary is important in determining the minimum mass ratio of W UMa
binaries. In addition, we investigate the mass loss during the
merging process of W UMa systems. We find that W UMa systems should
lose a large amount of mass to avoid these merged stars rotating
faster than the break-up velocities. If the effect of the magnetic
braking is considered, the angular momentum loss may be more
efficient and the mass loss would decrease.

\begin{table*}
\begin{footnotesize}
Table~1.\hspace{4pt} The physical parameters of W UMa binaries.\\
\begin{minipage}{16cm}
\begin{tabular}{l|cccccccccc}
\hline\hline\ {Stars}&{$q_{ph}$}&{$M_{1}$}&{$R_{1}$}&{$P$}&{$v_{\rm
e}$}&{$v_{\rm b}$}&{$R_{\rm ex}$}&
{${\rm \delta}M$}&{${\rm \delta}M_{\rm mb}$}&{References}\\
&&{($M_{\rm \odot}$)}&{($R_{\rm \odot}$)}&{(days)}&{(km s$^{-1}$)}&{(km s$^{-1}$)}
&{($R_{\rm \odot}$)}&{($M_{\rm \odot}$)}&{($M_{\rm \odot}$)}\\
\hline
AW UMa        &0.078 &1.79 &1.9  &0.4387&819.31&423.85&9.73&0.645&0.349&(1)\\
SX Crv        &0.0787&1.246&1.347&0.3166&804.60&419.98&6.67&0.439&0.238&(2)\\
V870 Ara      &0.082 &1.503&1.67 &0.3997&778.28&414.27&8.22&0.518&0.280&(3)\\
FP Boo        &0.096 &1.614&2.31 &0.6405&673.88&365.01&9.73&0.460&0.249&(4)\\
DN Bootis     &0.103 &1.428&1.71 &0.4476&709.84&399.05&7.59&0.439&0.238&(5)\\
CK Boo        &0.107 &1.442&1.521&0.3352&840.94&425.18&7.99&0.551&0.299&(4)\\
FG Hya        &0.111 &1.444&1.405&0.3278&792.26&442.69&6.96&0.514&0.279&(6)\\
GR Vir        &0.122 &1.376&1.49 &0.347 &788.40&419.64&7.34&0.492&0.266&(2)\\
$\epsilon$ CrA&0.128 &1.72 &2.12 &0.5914&655.84&393.33&8.69&0.487&0.264&(6)\\
DZ Psc        &0.135 &1.352&1.469&0.3661&730.89&418.92&6.71&0.444&0.241&(2)\\
V776 Cas      &0.138 &1.75 &1.821&0.4404&751.89&428.08&8.56&0.597&0.324&(7)\\
HN UMa        &0.140 &1.279&1.435&0.3825&681.55&412.26&6.11&0.385&0.209&(7)\\
V677 Cen      &0.142 &1.06 &1.19 &0.325 &664.60&412.13&4.94&0.309&0.168&(8)\\
V410 Aur      &0.144 &1.304&1.397&0.3663&691.20&421.89&6.04&0.402&0.218&(9)\\
AH Cnc        &0.149 &1.21 &1.36 &0.3605&681.99&411.89&5.80&0.368&0.199&(10)\\
TZ Boo        &0.153 &0.72 &0.97 &0.2976&587.95&376.22&3.56&0.180&0.098&(6)\\

\hline
\end{tabular}
\end{minipage}
\end{footnotesize}\\
{Columns: Stars-GCVS name of star; $q$-mass ratio; $M_1$-mass of the
primary; $R_1$-radius of the primary; $P$-orbital period; $v_{\rm
e}$, $v_{\rm b}$-the equatorial velocity and the break-up velocity
of the fast-rotating single star formed by the merger of W UMa
binary in the case of the conservation of mass and angular momentum;
$R_{\rm ex}$-the radius of the expanded fast-rotating single star
formed by the merger of W UMa binaries without mass loss; ${\rm
\delta}M$ and ${\rm \delta}M_{\rm mb}$-the lost mass during during
the merging process without and with considering the
magnetic breaking.\\
References in Table 1: (1) Pribulla et al. 1999; (2) Gazeas et al.
2005; (3) Szalai et al. 2007; (4) Gazeas et al. 2006; (5)
\c{S}enavc{\i} et al. 2008; (6) Yakut \& Eggleton 2005; (7) Zo{\l}a
et al. 2005; (8) Maceroni \& van¡¯tVeer 1996; (9) Yang, Qian \& Zhu
2005; (10) Zhang, Zhang \& Deng 2005.}
\end{table*}

\begin{figure}
\centerline{\psfig{figure=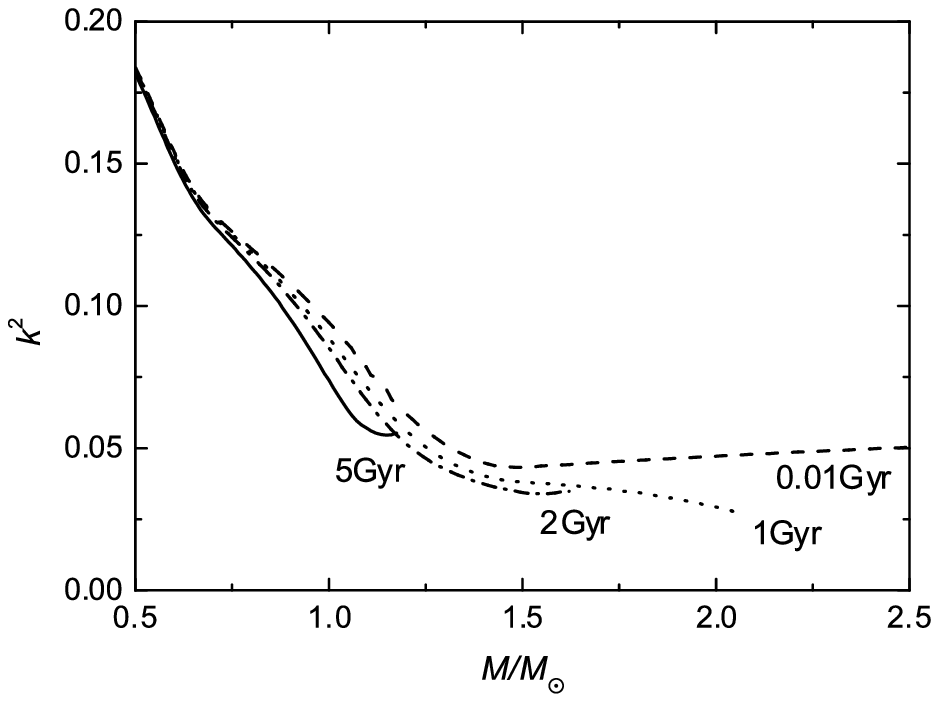,width=7.5cm}} \caption{The
relation of $k^2$ $vs$ $M$ for stars at age=10Myr, 1Gyr, 2Gyr, and
5Gyr, respectively.} \label{fig1}
\end{figure}

\begin{figure}
\centerline{\psfig{figure=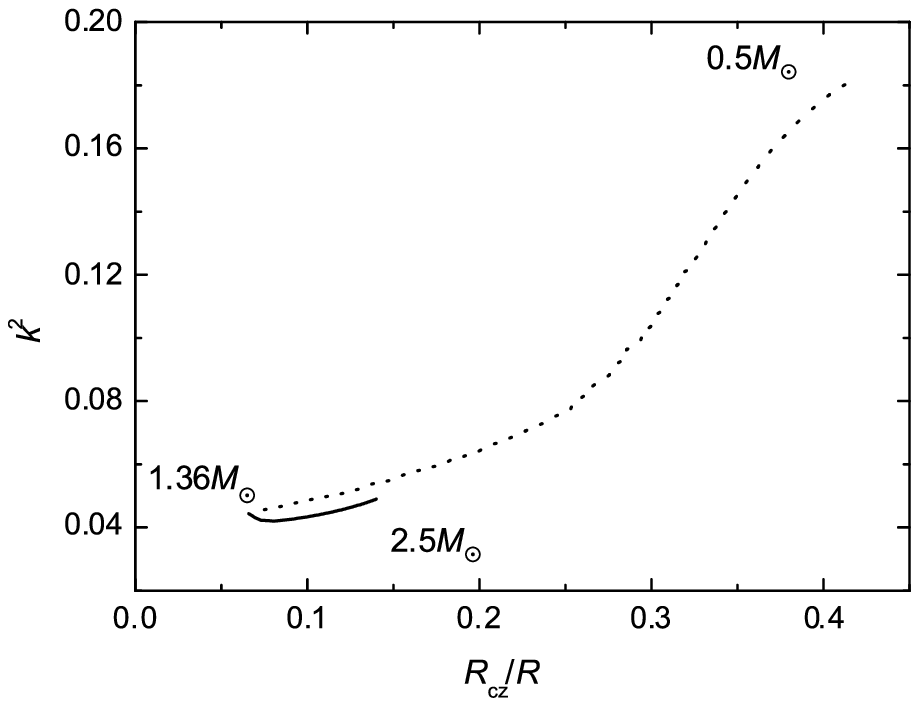,width=7.5cm}} \caption{The
relation of the fractional radius of the convection zone ($R_{\rm
cz}/R$) and $k^2$ for stars age=10Myr. The dotted line represents a
relation between $R_{\rm cz}/R$ and $k^2$ of the stars with a
convective envelope and the solid line represents the relation
between $R_{\rm cz}/R$ and $k^2$ of the stars with a convective
core. The change between convective envelope to core occurs at
1.36$M_{\rm \odot}$. }\label{fig2}
\end{figure}

\begin{figure}
\centerline{\psfig{figure=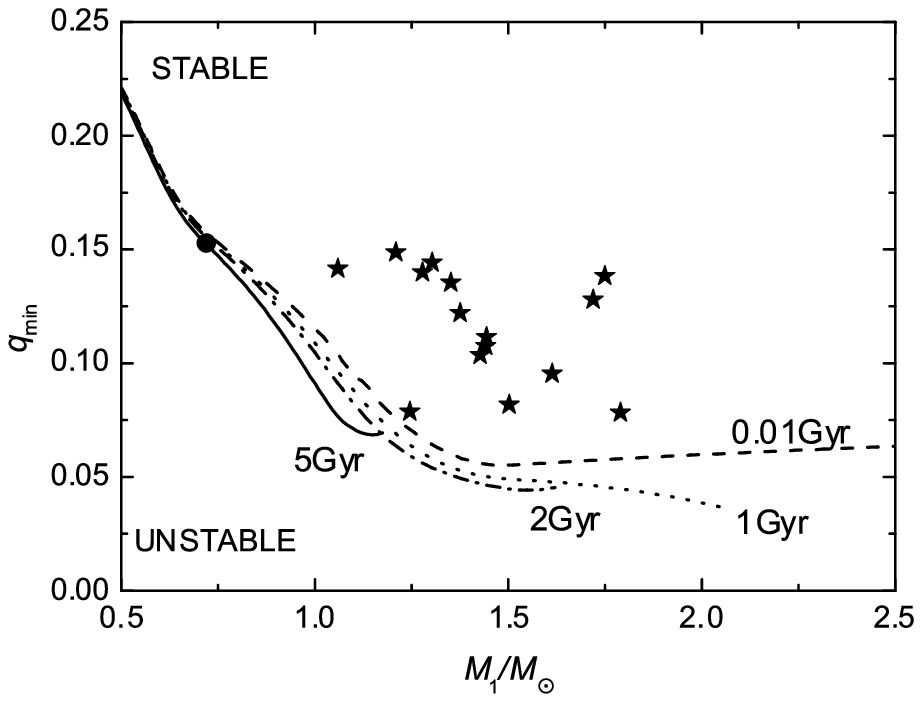,width=7.5cm}} \caption{The
relation of $q_{min}$ and the mass of the primary of W UMa binary.
Filled stars represent the observed W UMa binaries with extreme mass
ratios and filled circle represents TZ Boo.} \label{fig2}
\end{figure}

\section{the minimum mass ratio}
\citet{Rasio 1995} predicted that the dynamical stability limit
should depend on the structure of W UMa systems with given masses
$M_1$ and $M_2$. We use a stellar evolution code to determine the
dimensionless gyration radii ($k^2$) of stars with different masses
and ages, and study the effect of the interior structure (the
dimensionless gyration radius) of components with different masses
on the minimum mass ratio. This code is originally developed by
\citet{Eggleton 1971,Eggleton 1972} and \citet{Eggleton 1973} and
has been updated with latest input physics during the last three
decades \citep[e.g.][]{han94,pols95,pols98,Nelson 2001,Eggleton
2002}. The observations of W UMa binaries show a well defined
short-period limit of about 0.22 d \citep{ruc08}, which is
equivalent to a lower mass limit for the primary of approximately
0.6$M_{\rm \odot}$ \citep{ste06}. So we calculate the dimensionless
gyration radii $k$ of the stars with the solar metallicity (Z=0.02)
and with mass ($M$) between 0.5 and 2.5$M_{\rm \odot}$.

Fig. 1 shows a relation of $k^2$ $vs$ $M$ for the stars at age of
10Myr, 1Gyr, 2Gyr, and 5Gyr, respectively. It is seen from Fig.1
that $k^2$ decreases with increasing age and this can help explain
the existence of low-$q$ W UMa systems as suggested by \citet{Li
2006} and \citet{Arbutina 2007}. In addition, we also found that
$k^2$ decreases with increasing mass of the star if the star's mass
is less than about $1.3M_{\rm \odot}$, and above this mass $k^2$ is
roughly constant. This might be caused by the difference in the
structure of the main-sequence stars with different masses. We
displays the relation of the fractional radius of the convection
zone ($R_{\rm cz}/R$) and $k^2$ for stars age=10Myr in Fig. 2 (where
$R_{\rm cz}$ is the thickness of the convective zone and $R$ the
stellar radius). The dotted line represents a relation between
$R_{\rm cz}/R$ and $k^2$ of the stars with a convective envelope and
the solid line represents the relation between $R_{\rm cz}/R$ and
$k^2$ of the stars with a convective core. It is found that for the
stars with $M \leq 1.36M_{\rm \odot}$, the convection zone is closed
to the surface. If the mass increases from $0.5M_{\rm \odot}$ to
$1.36M_{\rm \odot}$, the fractional radius of the convection zone
($R_{\rm cz}/R$) decreases from 0.41 to 0.07 and the dimensionless
gyration radius decreases from 0.18 to 0.045 with decreasing $R_{\rm
cz}/R$. For the stars with $M \geq 1.36M_{\rm \odot}$, the
convection zone is closed to the center of star. The dimensionless
gyration radius of the star is at a stable value of about 0.047 as
$R_{\rm cz}/R$ increases from 0.066 to 0.14.

\citet{Li 2005} argued that the efficient energy transfer would
decrease the dimensionless gyration radius of the secondary ($k_2$)
in a W UMa system and the value of $k_2$ is not strongly different
from that of the dimensionless gyration radius of the primary
($k_1$), although there is a significant difference in the masses of
the primary and the secondary. In fact, the radius of the secondary
of each W UMa binary is much larger than that of a main-sequence
star with the same mass and its mass distribution has been greatly
changed compared with the main-sequence star due to energy transfer
\citep{Li 2005,Yakut 2005,Li08}. \citet{Rasio and Shapiro 1995}
found that the dynamical stability limit of W UMa systems is at a
contact degree ($F$) of about 70 per cent
($F=\frac{\Omega-\Omega_{\rm in}}{\Omega_{\rm out}-\Omega_{\rm
in}}\times 100\%$, where $\Omega$ is the surface potential of W UMa
system, $\Omega_{\rm in}$ and $\Omega_{\rm out}$ are the potentials
of the inner and the outer Lagrangian points, respectively).
Therefore, assuming that the dimensionless gyration radii for both
components are equal ($k_1^2$=$k_2^2$) and using the relation
between the minimum mass ratio and $k^2$ ($F$=0.7) given by
\citet{Li 2006}, we can obtain a relation between the theoretical
minimum mass ratio and the mass of the primaries for W UMa systems,
which is shown in Fig. 3. It is seen in Fig. 3 that the minimum mass
ratio decreases with the evolutionary degree of the W UMa systems as
suggested by \citet{Rasio 1995} and \citet{Li 2006}. This suggests
that the dynamical stability limit of W UMa systems indeed depends
on the evolutionary status of W UMa systems. It is also found that
the minimum mass ratio of the young W UMa binaries with an age of 10
Myr decreases with increasing mass of the primary if the primary's
mass is less than about 1.3$M_{\rm \odot}$, and above this mass the
ratio is roughly constant.

We collected the physical parameters of some W UMa systems (listed
in Table 1) with extreme mass ratios from the literature. The
observed systems are also plotted in Fig. 3 with filled stars. TZ
Boo is indicated with a different symbol (filled circle) because it
has been found to be a quadruple system and its spectra has been
contaminated by the third and fourth bodies \citep{Pribulla 2009}.
In addition, we do not include V857 Her, which has the lowest mass
ratio of $q$=0.065 \citep{Qian 2006}, in our analysis as the primary
mass is unknown. We predict its mass must be greater than
1.25$M_{\rm \odot}$. It is seen in Fig. 3 that W UMa systems are
above the theoretical curves except for TZ Boo, i.e. they are
located in the stable region although some of these systems have
mass ratio lower than the minimum mass ratio predicted by the
previous theory. TZ Boo located in the unstable region might be due
to the uncertainty of our stellar models that the metallicity effect
is not considered. Another probable reason is the presence of the
additional companion(s).

\section{the merger of W UMa binaries}
\citet{Li08} argued that the mass ratio of W UMa systems become
smaller and smaller owing to their dynamical evolution during their
evolution. Therefore, they would merge into fast rotating stars due
to Darwin's instability if their mass ratios have become smaller
than the cutoff mass ratio of W UMa systems. When a W UMa binary
begins to coalesce into a fast-rotating single star due to Darwin's
instability, its orbital angular momentum can be expressed as
$J_{\rm orb} = 3J_{\rm spin}$ \citep{Hut 1980, Eggleton 2001}. We
assumed that the W UMa binary is in synchronous rotation (i.e.
$\omega_{{\rm spin},1}= \omega_{{\rm spin},2}= \omega_{\rm orb} =
\omega_0=2\pi /P_{\rm orb}$, where $\omega_{{\rm spin},1}$ and
$\omega_{{\rm spin},2}$ are the spin angular velocities of two
components and $\omega_{\rm orb}$ is the orbital angular velocity of
the system). So the total angular momentum of a merging W UMa binary
can be approximately expressed as
\begin{equation}
J_{\rm b,tot}=J_{\rm orb}+ J_{\rm spin} \approx 4J_{\rm spin}
=4(k_1^2M_1R_1^2+k_2^2M_2R_2^2)\omega_0.
\end{equation}
where $M_{1,2}$ and $R_{1,2}$ are the masses and radii of the
primary and the secondary in solar units, and $k_{1,2}$ the
dimensionless gyration radii for both components. We assumed that
the dimensionless gyration radii for both components of W UMa binary
are equal ($k_1^2$=$k_2^2$) as showed by \citet{Li 2005}. When a W
UMa binary with a mass ratio of $q$ merges into a single star, we
have
\begin{equation}
J_{\rm
b,tot}=\frac{4(1+q^{1.92})}{1+q}k_1^2(M_1+M_2)R_1^2\omega_0,
\end{equation}
where $q$ ($=M_2/M_1$) is the mass ratio.

The angular momentum of the fast-rotating single star formed by
the merger of W UMa binary can be written as
\begin{equation}
J_{\rm s} = k^2MR^2\omega = k^2MRv_e,
\end{equation}
where $M$ and $R$ are the mass and radius of the fast-rotating
single star formed by merger of the W UMa system in solar units, $k$
the dimensionless gyration radius, $\omega$, $v_e$ the spin angular
velocity and the equatorial velocity of the fast-rotating star
formed from the merger.

\subsection{The merger of W UMa binaries without angular momentum loss}
We assumed that the merged star and the primary have the same radius
and the same dimensionless gyration radius ($k_1^2 \approx k^2$ and
$R_1 \approx R$) because the mass ratio of W UMa system is very low
at the beginning of merger. If the total angular momentum and total
mass are conserved in the course of the merger ($J_{\rm s}=J_{\rm
b,tot}$, $M=M_1+M_2$), using equation (2) and (3), we can get
\begin{equation}
\omega \approx \frac{4(1+q^{1.92})}{1+q}\omega_0,
\end{equation}
and
\begin{equation}
v_{\rm e} =R\omega \approx \frac{4(1+q^{1.92})}{1+q}R_1\omega_0.
\end{equation}
In addition, we can calculate the break-up velocities ($v_{\rm b}$)
of the single stars formed by merger of W UMa systems if the W UMa
systems with extreme mass ratios would merge into fast-rotating
stars. The break-up velocity ($v_{\rm b}$) of a single star formed
by the merger can be written as
\begin{equation}
v_{\rm b} \approx (\frac{GM_1}{R_1})^{\frac{1}{2}}.
\end{equation}
Based on equations (5) and (6), the equatorial velocities and the
break-up velocities for the single stars formed from the merger of W
UMa systems are determined, and they are listed in Table 1. It is
found that the distribution of the equatorial velocities of the
fast-rotating single stars ranges from 588 to 819 km s$^{-1}$. If
the angular momentum and mass of W UMa systems are conserved in the
course of merger, the single stars formed by the merger of W UMa
binaries rotate at a velocity faster than their break-up velocities.
This suggests that the applicability of the assumptions ($J_{\rm
s}=J_{\rm b,tot}, R_1\simeq R$) is unreasonable. In the course of
the merger, W UMa systems should lose a large amount of mass and
angular momentum, or the merged stars have expanded to a very large
radius compared with that of a main-sequence star with the same
mass.

It is noted that FK com stars are rapidly rotating G-type giants
which might result from the merger of close (W UMa) binaries
\citep{Bopp Rucinski 1981, Bopp Stencel 1981, Webbink 1976}. We
assumed that the fast-rotating star expands to a very large radius
($R=R_{ex} \gg R_1$) and $J_{\rm s}=J_{\rm b,tot}$. Using equation
(2) and equation (3), we can calculate the radius of the expanded
merged star formed by the merger of W UMa binary
\begin{equation}
R_{\rm ex} \approx
\frac{4(1+q^{1.92})}{1+q}\frac{R_1^2\omega_0}{v_e}.
\end{equation}
It is found that the radius of the expended single star depends on
the equatorial velocity. \citet{Rucinski 1990} showed that the
rotational velocity ($v \, \sin i$) of FK Com star is about 160km
s$^{-1}$. So we assumed that the equatorial velocity of the expanded
merged star is 160 km s$^{-1}$. Then we found that the expanded
merged stars formed by the merger of W UMa binaries have the radii
of about 3.6$\sim$9.7$R_{\rm \odot}$ (listed in table 1) and are
3.7$\sim$5.3 times the radii of the primaries.

\subsection{The mass loss without the magnetic braking}
If the merged stars do not expand, W UMa systems should lose a large
amount of mass and angular momentum in the course of the merger. The
mass loss (${\rm \delta}M$) during the merging process can be
calculated by using the following formulae:
\begin{equation}
J_{\rm b,tot}=k_{\rm s}^2(M_1+M_2-{\rm \delta}M)R_{\rm s}v_{\rm
s}+{\rm \delta}MR_{\rm esc}v_{\rm esc},
\end{equation}
where $k_{\rm s}$, $R_{\rm s}$ and $v_{\rm s}$ are the dimensionless
gyration radius, the radius and the equatorial velocity of the
merged star; $R_{\rm esc}$ and $v_{\rm esc}$ are the radius and the
velocity of the stellar wind escaped from the system. We assumed
that the merged star and the primary have the same radius and the
same dimensionless gyration radius ($k_1^2 \approx k_{\rm s}^2$ and
$R_1 \approx R_{\rm s}$). \citet{De Marco 2005} gave a mean value of
the rotational velocity for five fast rotating blue stragglers to be
of about 160 km s$^{-1}$. So we assumed that the equatorial velocity
of the merged star is $v_{\rm re} \approx 160$km s$^{-1}$. If the
effect of magnetic braking is not considered, we can take that
$R_{\rm esc} \approx R_{\rm s} \approx R_1$ and $v_{\rm esc} \approx
v_{\rm s} \approx 160$km s$^{-1}$. Then, using equations (2) and
(8), we obtain
\begin{equation}
{\rm
\delta}M=[\frac{4(1+q^{1.92})}{1+q}\frac{R_1\omega_0}{160}-1]\frac{k^2}{1-k^2}(M_1+M_2).
\end{equation}
We assumed $k^2$ to be about 0.075. Based on equation (9), the lost
mass during the merging process of W UMa systems can be calculated
and listed in Table 1. It is found that the distribution of the mass
loss has a range from $0.18\sim0.65M_{\rm \odot}$, which is about
$21\sim33$ per cent of the total mass.

\subsection{The mass loss with the magnetic braking}
The stars which have convective envelope can be magnetically braked
and slowed down faster than just angular momentum loss in stellar
wind \citep{Tout 1992}. The stellar wind is forced by the magnetic
field to corotate out to the Alfv\'{e}n radius ($R_A$) then escapes
freely ($R_{\rm esc}=R_A$). Then the same mass lost from the systems
would take away more angular momentum than that escaped from the
surface of the systems. We studied the mass loss (${\rm
\delta}M_{mb}$) in the merging process of W UMa binary when the
effect of magnetic braking is included. We assumed that the stellar
wind is in synchronous rotation with the merged stars ($v_{\rm
esc}/R_A=v_{\rm s}/R_{\rm s}$) and $v_{\rm s} \simeq 160$km
s$^{-1}$. Then, the mass loss can be written as
\begin{equation}
{\rm
\delta}M_{mb}=[\frac{4(1+q^{1.92})}{1+q}\frac{R_1\omega_0}{160}-1]\frac{k^2}{(\frac{R_A}{R_1})^2-k^2}(M_1+M_2).
\end{equation}
The Alfv\'{e}n radius can be expressed as \citep{Tout 1992}:
\begin{equation}
\frac{R_A}{R_1}=1.1f^{-1/4}(\gamma/10^{-2})^{1/2},
\end{equation}
in which
\begin{equation}
f=\omega/\omega_b,
\end{equation}
where $\gamma$ is the efficiency of dynamo regeneration
($\sim10^{-2}$), $\omega$ and $\omega_b$ are the angular velocity
and the break-up angular velocity of the star. The break-up
angular velocity reads
\begin{equation}
\omega_b= (\frac{GM_1}{R_1^3})^{\frac{1}{2}},
\end{equation}
and the angular velocity reads
\begin{equation}
\omega =
\frac{G^{\frac{1}{2}}(M_1+M_2)^{\frac{1}{2}}}{A^{\frac{3}{2}}}.
\end{equation}
Based on equations (12), (13) and (14), we can obtain
\begin{equation}
f=(1+q)^{\frac{1}{2}}(\frac{0.49q^{-2/3}}{0.6q^{-2/3}+{\rm
ln}(1+q^{-1/3})})^{\frac{3}{2}}.
\end{equation}
If $q \sim 0.1$, we can obtain
\begin{equation}
f \approx 0.4611.
\end{equation}
From equation (11) and equation (16), we find
\begin{equation}
\frac{R_A}{R_1}\approx1.335.
\end{equation}
Hence, the mass loss (${\rm \delta}M_{mb}$) is determined by using
equation (10) and equation (17) with considering the effect of
magnetic braking and listed in Table 2. The distribution of the mass
loss has a range from $0.10\sim0.35M_{\rm \odot}$ in the course of
merger which is about $12\sim18$ per cent of the total masses.

\section{Discussion and conclusions}
In this paper, we investigated the minimum mass ratio of W UMa
systems that have different primary masses. In addition, we studied
the mass loss during the merger of W UMa systems.

\citet{Arbutina 2009} has investigated the theoretical stability
limit of W UMa binary, where he has considered the effects of
rotation which would increase the central concentration. He gave the
minimum mass ratio of W UMa binaries to be 0.070-0.074. Considering
the different structure of the primaries, we found that the minimum
mass ratio of the young W UMa binaries with an age of 10 Myr
decreases with increasing mass of the primary if the primary's mass
is less than about 1.3$M_{\rm \odot}$, and above this mass the ratio
is roughly constant. The minimum mass ratio of binaries with $M_1
\geq 1.3M_{\rm \odot}$ is lower than the value given by
\citet{Arbutina 2009}. This is mainly because he considered the
effect of rotation based on $n=3$ polytrope (which has $k^2 \approx
0.075$). By comparing the theoretical minimum mass ratio with the
observational data, it is found that the observational systems are
in the stable region except for TZ Boo. This means that these
observed W UMa systems are dynamically stable and the existence of
low-$q$ systems can be explained by the different structure of the
primaries with different masses. This suggests that the
dimensionless gyration radius and thus the structure of the primary
is very important in determining the minimum mass ratio. A W UMa
system with a less massive primary will merge at a larger mass
ratio. Therefore, it is necessary to consider the different
structure of the primaries for the study of the dynamical stability
of W UMa systems.

Previous theoretical studies have argued that W UMa binary would
eventually merge into a single star \citep{Webbink 1976, Webbink
1985, Tutukov 1987, Mateo 1990}. Assuming that the angular momentum
and mass are not lost from W UMa systems during the merging process,
it is found that the merged stars rotate faster than their break-up
velocities, which is unreasonable. One possible explanation is the
merged stars expand to a very large radius which is about be
3.7$\sim$5.3 times the radii of the primaries. These expanded merged
stars can be observed like FK com-type stars. We needs to obtain the
parameters of FK Com-type stars and make comparisons with our
results.

Another possible explanation is that during the merger, W UMa
systems should lose a large amount of mass and angular momentum.
\citet{Chen 2008} predicted that a large amount of mass
($\sim0.5M_{\rm \odot}$) must be lost from W UMa systems in the
course of merger if the theoretical model can match the
observations. However, they did not give a physical mechanism for
the mass loss. We calculated that the mass loss during the merging
process of W UMa system would be of 21$\sim$33 per cent of the total
mass. If the effect of magnetic braking is considered, the angular
momentum loss due to mass loss may be more efficient and the mass
loss would decrease to be 12$\sim$18 per cent of their total masses.
Our results are smaller than that predicted by \citet{Chen 2008}.
This might be due to the lack of W UMa systems with $M_1 \geq
2.0M_{\rm \odot}$ in our sample and these systems would lose more
mass during the merging process.

There is a significant difference between the theoretical prediction
and the observations of the rotation velocities of the blue
stragglers and FK Com-type giants. In the color-magnitude diagrams
of globular clusters, some W UMa systems are, in fact, in the region
of the blue stragglers, and there are 20 W UMa binaries observed
among about 900 blue stragglers \citep[][and references
therein]{von02,von03,ruc00}. This implies that the formation of some
blue stragglers is related to the merger of W UMa systems. In
addition, \citet{Taam 2000} suggested that binaries in the common
envelope phase at short orbital periods must eventually merge into a
single star since the massive component cannot be on the giant
branch. Our study indicates that the significant angular momentum
and mass might be lost from W UMa system in the course of the
merging process, and this kind of mass and angular momentum loss
might be driven by the release of orbital energy of the secondaries.
This is similar to common-envelope evolution. When the secondary of
W UMa system with high spin velocity merges into the primary, the
orbital energy is deposited into the envelope, disrupting it. This
energy can only make some mass of the envelope to be ejected,
because for the stars near the main sequence, the binding energy of
the envelope is too large for energy from the orbital motion to
completely eject it \citep{Taam 2000}. The other possible mechanism
are first a circum-stellar disk. The mass lost from the system might
form a circum-stellar disk since the the equatorial rotational
velocity is the fastest. This circum-stellar disk would extract the
angular momentum from the single stars formed by the merger of the W
UMa systems through the tidal torque \citep{Chen 2006}. Second, the
angular momentum loss is caused by W UMa systems' companion(s)
through tidal friction. The progenitors (W UMa binaries) of some
blue stragglers and FK Com-type giants might be formed in primordial
and dynamically formed triple systems \citep{Leonard 1996, Perets
2009}. \citet{Pribulla 2006} showed that most W UMa binaries exist
in multiple systems. The presence of distant companions can
facilitate not only the formation of W UMa binaries but also the
deceleration of the fast-rotating single stars formed by the merger
of W UMa binaries. Third, for FK Com-type giants, their radii are
much larger than that of the main-sequence stars with the same
masses and their dimensionless gyration radii have risen to a higher
value because their envelopes have become convective. It would
undertake the decrease in the rotation velocities of FK Com-type
giants.

\section*{ACKNOWLEDGEMENTS}
It is a pleasure to thank an anonymous referee for his/her many
suggestions and comments which considerably improved the paper. This
work was partly supported by the Chinese Natural Science Foundation
(10673029, 10773026, 10521001, 2007CB815406 and 10821061), the
Foundation of Chinese Academy of Sciences (KJCX2-YW-T24) and by the
Yunnan Natural Science Foundation (2007A113M).

\bsp

\label{lastpage}

\end{document}